\documentclass[11pt,a4paper]{article}
\usepackage{jcappub}

\usepackage{graphicx,epsfig,amssymb,times} 
\usepackage{amsmath,amsfonts}
\usepackage{bm}
\usepackage{epstopdf}
\usepackage[caption=false]{subfig}
\usepackage[usenames]{color}     
\usepackage{tikz}
\usetikzlibrary{decorations.pathmorphing}
\usepackage{float}

\usepackage{multirow}
\usepackage{lipsum}  
\usepackage{float}
\usepackage{color}
\usepackage{cleveref}

\usepackage{comment}

\usepackage{parskip} 
\usepackage{indentfirst}
\definecolor{coolblack}{rgb}{0.0, 0.18, 0.39}
\definecolor{darkred}{rgb}{0.5,0,0}
\definecolor{darkgreen}{rgb}{0,0.5,0}
\definecolor{darkblue}{rgb}{0,0,0.5}
\definecolor{lapislazuli}{rgb}{0.15, 0.38, 0.61}
\definecolor{venetianred}{rgb}{0.78, 0.03, 0.08}
\definecolor{bleudefrance}{rgb}{0.19, 0.55, 0.91}
\definecolor{dogwoodrose}{rgb}{0.84, 0.09, 0.41}

\usepackage{wrapfig}
\usepackage{parskip} 
\usepackage{indentfirst}
\usepackage[linktocpage=true]{hyperref}
\hypersetup{
    colorlinks,
    linkcolor={red!50!black},
    citecolor={blue!50!black},
    urlcolor={blue!80!black}
}

\usepackage{mathtools}
\usepackage{amssymb}
\usepackage{mathrsfs}
\usepackage{amsmath}
\usepackage{xcolor}
\usepackage{physics}

\newcommand{\eg}{{\it e.g. }}

\newcommand{\cR}{\ensuremath\mathcal{R}}

\newcommand{\lr}[1]{\left(#1\right)}

\newcommand{\beq}{\begin{equation}}
\newcommand{\eeq}{\end{equation}}

\newcommand{\rhob}{\rho_\textrm{b}}
\newcommand{\rhoc}{\rho_\textrm{c}}

\definecolor{darkred}{rgb}{0.55, 0.0, 0.0}
\definecolor{darkmagenta}{rgb}{0.55, 0.0, 0.55}
\definecolor{lincolngreen}{rgb}{0.11, 0.35, 0.02}
\definecolor{mygreen}{rgb}{0.05, 0.35, 0.1}

\newcommand{\rc}{\mathcal{R}_\textrm{c}}
\newcommand{\lp}{\ell_{\mathrm{Pl}}}

\definecolor{darkgreen}{rgb}{0.2,0.45,0.2}

\definecolor{darkred}{rgb}{0.5,0,0}
\definecolor{venetianred}{rgb}{0.78, 0.03, 0.08}

\title{A diffeomorphism invariant family of metric-affine actions for loop cosmologies}

\author[a]{Adri\`a Delhom,}
\author[b,c]{Gonzalo J. Olmo,}
\author[a]{and Parampreet Singh}

\affiliation[a]{Department  of  Physics  and  Astronomy, Louisiana  State  University,  Baton  Rouge,  LA  70803,  USA}
\affiliation[b]{Departamento de F\'isica Te\'orica and IFIC, 
Centro Mixto Universidad de Valencia - CSIC. 
Universidad de Valencia, Burjassot-46100, 
Valencia, Spain}
\affiliation[c]{Universidade Federal do Cear\'a (UFC), Departamento de F\'isica,\\ Campus do Pici, Fortaleza - CE, C.P. 6030, 60455-760 - Brazil.}

\emailAdd{adria.delhom@gmail.com}
\emailAdd{gonzalo.olmo@uv.es}
\emailAdd{psingh@lsu.edu}

\abstract{
In loop quantum cosmology (LQC) the big bang singularity is generically resolved by a big bounce. This feature holds even when modified quantization prescriptions of the Hamiltonian constraint are used such as in mLQC-I and mLQC-II. While the later describes an effective description qualitatively similar to that of standard LQC, the former describes an asymmetric evolution with an emergent Planckian de-Sitter pre-bounce phase even in the absence of a potential. We consider the potential relation of these canonically quantized non-singular models with effective actions based on a geometric description. We find a 3-parameter family of metric-affine $f(\cR)$ theories which accurately approximate the effective dynamics of LQC and mLQC-II in all regimes and mLQC-I in the post-bounce phase. Two of the parameters are fixed by enforcing equivalence at the bounce, and the background evolution of the relevant observables can be fitted with only one free parameter. It is seen that the non-perturbative effects of these loop cosmologies are universally encoded by a logarithmic correction that only depends on the bounce curvature of the model. In addition, we find that the best fit value of the free parameter can be very approximately written in terms of fundamental parameters of the underlying quantum description for the three models. The values of the best fits can be written in terms of the bounce density in a simple manner, and the values for each model are related to one another by a proportionality relation involving only the Barbero-Immirzi parameter.
}


\begin{document}

\date{\today}

\maketitle

\section{Introduction}\label{sec:intro}

The ubiquitous presence of classical singularities in General Relativity (GR) \cite{Penrose:1969pc,Hawking:1970zqf,Senovilla:2014gza} suggests that it cannot be understood as the ultimate description of the gravitational interaction. Indeed, if gravity is to be quantized, it is well known that, due to its non-renormalizability, GR cannot be seen as a fundamental theory of gravitation but must rather be interpreted as an effective field theory (EFT) with a cutoff at the Planck scale \cite{Donoghue:1994dn,Burgess:2003jk,Donoghue:2012zc}. One can thus try to solve the problem of singularities within the EFT paradigm, by looking for speccific effective theories that do not contain singular solutions for physically sensible initial data. This was the approach followed in \cite{Brandenberger:1993ef}, where higher-order curvature effective theories were found such that all homogeneous and isotropic solutions are non-singular within. The hope is that such kind of theories may shed some light into the properties of the underlying quantum gravity description although, to that end, one has to go beyond the EFT paradigm and understand what possible UV completions these effective theories can have.

The (in principle) most straightforward way to aim for a fundamental theory of gravity is to seek for a renormalizable action for the gravitational interaction that recovers GR in the regimes where it has been tested. Lovelock's theorem ensures that new degrees of freedom will propagate unless we depart from metricity and/or diffeomorphism invariance (namely background independence) if we stick to four space-time dimensions. These new degrees of freedom usually generate ghost instabilities, and standard quantizations of such theories suffer from a breakdown of unitarity \cite{Donoghue:2012zc}, although non-standard quantizations in terms of an anti-linear Hamiltonian (rather than hermitian) have been suggested as a possible solution to such problem \cite{Bender:2007wu,Mannheim:2021xrg,Mannheim:2020ryw,Mannheim:2018ljq}. Other directions that have been taken in the quest for a fundamental theory of gravity include going beyond perturbative modifications of the Einstein-Hilbert action.

In the past decades, much theoretical effort has been dedicated to find candidates for quantum theories of gravity that can be interpreted as fundamental, one of the typical requirements for these candidates being the ability to tame the classical singularities. One of such approaches is Loop Quantum Gravity (LQG), a non-perturbative canonical quantization of GR where the canonical phase space coordinates {{of the gravitational sector are the Ashtekar-Barbero variables -- the holonomies of the connection and the fluxes associated with the triads.}} While the dynamics of the full theory remains to be uncovered, much progress has been made in implementing the quantization techniques of LQG to  symmetric scenarios of physical relevance, such as cosmological \cite{Ashtekar:2011ni} and black hole space-times \cite{Ashtekar:2023cod}. The idea behind these approaches is to reduce the dimensionality of the phase space by enforcing the constraints provided by the {{underlying symmetries of the spacetime}}, and then quantize the degrees of freedom of the resulting minisuperspace by using loop quantization techniques. {{As a result,  the classical differential geometry of Einsteinian gravity is replaced by a discrete quantum geometry. Extensive investigations in loop quantization of cosmological models show that this underlying quantum geometry results in a bounce which replaces the big bang \cite{Ashtekar:2006rx, Ashtekar:2006wn, Ashtekar:2007em}, with similar results on the resolution of the central singularity in black hole spacetimes (see eg. \cite{Ashtekar:2018lag}). The dynamical resolution of the singularity, which occurs when the energy density reaches the Planckian regime,  manifests itself even at the level of consistent probabilities. Unlike in the Wheeler-DeWitt theory \cite{Craig:2010vf}, the probability for a universe to encounter a singularity is zero, and to have a bounce is unity \cite{Craig:2013mga}.

Interestingly, one can obtain an effective spacetime description in LQC \cite{Taveras:2008ke} which captures the underlying quantum evolution extremely well \cite{Ashtekar:2006wn}. Numerical simulations with high performance computing  confirm this feature \cite{Diener:2014mia}, including in presence of anisotropies \cite{Diener:2017lde, Singh:2018rwa}. These investigations show that for a wide variety of states, bounce in LQC is a robust phenomenon \cite{Diener:2014hba}.  Assuming the validity of the effective dynamics, one can prove that for arbitrary matter there is a generic resolution of singularities in LQC for isotropic as well as anisotropic spacetimes \cite{Singh:2009mz, Singh:2011gp, Singh:2014fsy, Saini:2016vgo, Saini:2017ipg, Saini:2017ggt}. Our manuscript will assume the validity of the effective spacetime description for all regimes and all of the loop cosmology models considered.

As in any approach to quantum gravity, the application of loop quantization techniques also faces inherent ambiguities in the quantization procedure. One of these ambiguities results from the treatment of the Hamiltonian constraint, which has Euclidean and Lorentzian terms. In the spatially-flat models, using the symmetries of the space-time these terms can be combined before quantization resulting in the standard LQC \cite{Ashtekar:2003hd}. However, such a treatment is not applicable in general and in full LQG one must quantize both  terms independently. As a result, 
following Thiemann's regularization of the Hamiltonian constraint \cite{Thiemann:1996aw},  different choices to solve these ambiguities have provided two modified versions of quantum cosmologies \cite{Yang:2009fp,Dapor:2017rwv,Li:2018opr, Li:2018fco,Assanioussi:2019iye, Li:2021mop}, which following Ref. \cite{Li:2018opr} will be referred  to as mLQC-I and mLQC-II, and all 
 above models, including standard LQC
 as loop cosmologies. Note that as in standard LQC, one finds that there is a generic resolution of singularities in mLQC-I and mLQC-II \cite{Saini:2018tto}. 
 }}

{{
A natural question that permeates the LQG framework is how does one recover the spacetime covariance in the continuum spacetime description. }} From a classical perspective, this question can be reformulated for the different loop cosmologies by asking whether there exist any diff invariant theories that are able to provide an effective classical description that reproduces their background evolution and also the dynamics of their perturbations. At the background level, this question was already answered for {{standard}} LQC in \cite{Olmo:2008nf}, where a diff invariant metric-affine action that reproduced the standard LQC evolution was found. A natural question that arises after that finding, and the introduction of other loop cosmology models like mLQC-I and mLQC-II, is whether the different choices related to the quantization ambiguities affect this result, or also do yield an effective evolution that can be described by a diff invariant theory. In the affirmative case, a further direction to explore would be how different are the resulting effective classical theories for each of the models, or whether they can be embedded in a family of diff invariant actions where a choice of parameters leads to the different loop cosmology models. 

The exploration of alternative gravity theories in the last two decades offers multiple avenues in order to find diff invariant actions able to capture the dynamics of loop cosmologies. In this sense, an important piece of information that follows from LQC, mLQC-I, and mLQC-II is that the cosmic dynamics satisfies second-order equations {} in which the matter sources act nonlinearly, causing in that way nonperturbative effects (a bounce) without adding new dynamical degrees of freedom. This fact was already used in \cite{Olmo:2008nf} to identify metric-affine theories of the $f(\cR)$ type as a natural candidate for the effective description of loop cosmologies. 

In the metric-affine framework, one assumes that classical gravitation is not described only by a metric, but also by an independent affine or spin connection.\footnote{In principle, given an affine connection there is a canonical spin connection associated to it and viceversa. However, while researchers that view metric-affine gravity theories necessarily as gauge theories of gravity view the spin connection as the fundamental object; and curvature, torsion, and nonmetricity as field-strengths; other researchers see the affine connection as fundamental and regard torsion and nonmetricity just as pieces of this connection with certain geometrical meaning. For a more detailed and up to date account of the details of the metric-affine framework from different perspectives, see e.g. \cite{JimenezCano:2021rlu,Delhom:2021bvq}.}
Due to the independence between metric and  connection, curvature invariants do not contain higher-order derivatives of the metric, and for that reason metric-affine theories have been proposed in the past as a natural framework where  renormalizable ghost-free candidates could be found. It has been recently shown, however, that generic metric-affine theories are also swarmed by ghostly instabilities unless certain restrictive sub-classes or fine-tuned Lagrangians are considered \cite{BeltranJimenez:2019acz,Jimenez:2020dpn,Aoki:2019rvi,Percacci:2019hxn,Delhom:2021bvq,Jimenez-Cano:2022sds}. One of the ghost-free sub-classes are projective-invariant Ricci-Based Gravity (RBG) theories  where the Lagrangian is a generic function of the metric and the symmetrized Ricci tensor of the affine connection (see e.g. \cite{BeltranJimenez:2017doy,Delhom:2021bvq,Olmo:2022rhf}. Indeed, these theories have been shown to be dynamically equivalent to GR coupled to a non-linearly modified matter sector, so that they do not propagate extra degrees of freedom and naturally generate bouncing cosmologies, wormholes, and other exotic compact objects \cite{Afonso:2018bpv,Afonso:2018hyj,Afonso:2018mxn,Delhom:2019zrb,Delhom:2021bvq}. The simplest type of action that we can find within this sub-class is the metric-affine $f(\cR)$ family of theories, which we already mentioned in relation with the evolution in LQC \cite{Olmo:2008nf}. Beyond the $f(\cR)$ family and still ghost-free, one finds the RBG family, which includes quadratic gravity \cite{Olmo:2009xy,Olmo:2012nx} and the EiBI model \cite{BeltranJimenez:2017doy}, among others (see, for instance, \cite{Afonso:2021aho,Boudet:2022wmb,Olmo:2022ops}). 

In this work, we will assess the question of whether the effective continuum description that provides evolution\footnote{{In this work we will focus only in background evolution, leave the quest for equivalence for the dynamics of perturbations for future work.}} in mLQC-I and mLQC-II when coupled to perfect fluid with $\omega=1$ can also be derived from a covariant effective action within the $f(\cR)$ class coupled to the same fluid. We will show that there exists a common family of metric-affine $f(\cR)$ actions that faithfully describe the evolution predicted by LQC and mLQC-II, as well as the post-bounce phase of mLQC-I, and which reduce to the one found in \cite{Olmo:2008nf} for a particular choice of parameters. Interestingly, we find that the non-perturvative features of the evolution of the energy density with respect to the curvature are universally captured by a generic logarithmic correction $\ln(\cR/\cR_\mathrm{bounce})$ in the effective Lagrangian. Furthermore, once the general form of the family of effective $f(\cR)$ theories is specified, we will be able to write the values of the parameters leading to each of the considered loop cosmologies in terms of fundamental constants of the underlying LQC theory, {concretely the Barbero-Immirzi parameter $\gamma$ and the critical density at which the bounce occurs. In standard LQC this bounce density is given by $ \rho_\mathrm{b} = \rhoc= 3/8 \pi G \gamma^2 \Delta$ where $\Delta = 4 \sqrt{3} \pi \gamma \lp^2$ is the minimum area gap in quantum geometry. In mLQC-I and mLQC-II the bounce density is related to $\rho_c$ as discussed later.}Let us emphasize that, as explained above, this family of theories does not propagate extra degrees of freedom, is free from ghostly instabilities, and its dynamics can be mapped into GR coupled to a nonlinear matter sector. 

Other attempts to derive the effective dynamics of loop cosmologies already exist in the literature \cite{Ribeiro:2021gds,Miranda:2021oig}. However, they relay on an {\it order reduction} method that basically boils down to a perturbative approach designed to ignore the extra dynamical degrees of freedom of the considered theories. Such approaches, which somehow try to capture intrinsically nonperturbative phenomena within a perturbative approach, do not guarantee energy conservation (they break the Bianchi identities associated to diffeomorphism invariance), and are not robust under small perturbations, which excite the artificially frozen degrees of freedom. On the contrary, metric-affine $f(R)$ theories (and, more generally, RBGs) satisfy by construction the Bianchi identities, have no new dynamical degrees of freedom which may be excited by small perturbations, and generically exhibit nonperturbative effects \cite{Olmo:2022rhf}.

The structure of the paper is the following. In section \ref{sec:BackgroundEvolution}, we will briefly review the main features of the cosmological evolution for the different loop cosmology models and for metric-affine $f(\cR)$ theories, showing how the conditions for which both are equivalent at the physical level lead to a non-linear ODE for the metric-affine $f(\cR)$ Lagrangian. {{We will provide this description in cosmic time.}} In section \ref{sec:NumSol}, we will discuss the sensitivity of the solutions to small changes of at low curvatures, and physically meaningful numerical solutions for this ODE will be obtained. In section \ref{sec:Analytics}, we will find a family of metric-affine $f(\cR)$ Lagrangians that approximate the numerical solutions to the three ODEs corresponding to LQC, mLQC-I and mLQC-II to a high degree of accuracy, reducing to the known one for LQC found in \cite{Olmo:2008nf}. We will finish with a discussion on the results and future directions to be pursued.

\section{Background evolution equivalence}\label{sec:BackgroundEvolution}
In this section we will establish the conditions under which the evolution dictated by the effective theory is physically equivalent to the one described by the corresponding loop cosmological model by focusing in the relevant physical observables. To that end, we will first provide ingredients to describe evolution in both frameworks, the loop cosmology and the explicitly covariant, and then discuss the conditions for their equivalence.

\subsection{Evolution in loop cosmologies}\label{sec:BckEvolLQC}

The evolution of loop cosmology models can be expressed in terms of a modified Friedmann equation which modifies the GR evolution at energy densities close to the bounce, while quickly recovering GR away from the bounce scale. This modified Friedmann dynamics is obtained by quantizing the only gravitational degree of freedom that is left after a symmetry reduction that enforces homogeneity and isotropy in the phase space, namely the scale factor. This quantisation is carried out {\it \'a la Loop}, namely by rewriting GR in terms of Ashtekar-Barbero variables and using Dirac's method to quantise constrained theories. In isotropic and homogeneous backgrounds, of all the constraints in GR, only the Hamiltonian one is relevant, and it can be written as
\beq
\mathcal{H}_\mathrm{grav}=\mathcal{H}^{(E)}_\mathrm{grav}-(1+\gamma^2)\mathcal{H}^{(L)}_\mathrm{grav}\,,
\eeq
where $\mathcal{H}^{(E)}_\mathrm{grav}$ and $\mathcal{H}^{(L)}_\mathrm{grav}$ are the Euclidean and Lorentzian parts respectively  \cite{Yang:2009fp}, and $\gamma$ is the Barbero-Immirzi parameter, whose value can be fixed by black hole entropy computations in LQG to be $\gamma\approx0.2375$. {{As already mentioned, the quantization process has some ambiguities related to several possible treatments that can be given to the Lorentzian part of the constraint before quantising. These ambiguities lead to different effective Hamiltonians to describe cosmological evolution, leading to different Loop Cosmology models described by their respective modified Friedmann equations. In this paper, we consider two models that stem from these quantization ambiguities concerning how one treats the Euclidean and Lorentzian terms, which results in  mLQC-I and mLQC-II. Detailed investigations of their properties show that though the mLQC-II evolution is consistent with a symmetric quantum bounce as in standard LQC, this is not the case for mLQC-I, which has an asymmetric bounce for a consistent evolution \cite{Li:2018opr, Li:2021mop}. A striking difference between LQC, mLQC-II and mLQC-I is that the latter, unlike the other two quantizations, does severely forbid a cyclic evolution in cosmic time {even in the absence of a potential} \cite{Li:2021fmu}. }}

When coupled to a perfect fluid with a generic equation of state, the Friedmann equation for LQC reads
\begin{equation}
3H_{\rm LQC}^2=\kappa\rho\lr{1-\frac{\rho}{\rhob}},
\label{eq:LQCFriedmann}
\end{equation}
where $\rhob = \rhoc$ is the value of the energy density at the bounce, while the modified loop cosmology models are described respectively by the following modified Friedmann equations
\begin{equation}
3H^2_{\rm II}=\kappa \rho\left(1+\gamma^{2} \frac{\rho}{\rho_{c}}\right)\left(1-\frac{\left(\gamma^{2}+1\right) \rho / \rho_{c}}{\left(1+\sqrt{\gamma^{2} \rho / \rho_{c}+1}\right)^{2}}\right)
\label{eq:mLQC-IIFriedmann}
\end{equation}
for mLQC-II evolution \cite{Li:2018fco}, and
{{
\begin{equation}
3H^2_{\rm I+}=\kappa \alpha \rho_{\Lambda}\left(1-\frac{\rho}{\rhob^{\rm{I}}}\right)\left[1+\frac{1-2\gamma^2+\sqrt{1-\rho/\rhob^{\rm{I}}}}{4\gamma^2(1+\sqrt {1-\rho/\rhob^{\rm{I}}})}\frac{\rho}{\rhob^{\rm{I}}}\right],
\label{eq:mLQC-IpreFriedmann}
\end{equation}

\begin{equation}
3H^2_{\rm I-}=\kappa{{\rho}}\left(1-\frac{\rho}{\rhob^{\rm I}}\right)\left[1+\frac{\gamma^2}{1+\gamma^2}\left(\frac{\sqrt{\rho/\rhob^{\rm I}}}{1+\sqrt{1-\rho/\rhob^{\rm I}}}\right)^2\right],
\label{eq:mLQC-IFriedmann}
\end{equation}
}}
for pre-bounce and post-bounce phases of mLQC-I evolution respectively. Here $\rhob^\mathrm{I}=\rhoc/(4(1+\gamma^2))$ is the mLQC-I bounce density, and we can define analogously $\rhob^\mathrm{II}=4(1+\gamma^2)\rhoc$ which gives the mLQC-II bounce density.

\subsection{Evolution in $f(\cR)$}

In the Metric-affine formalism, the fundamental fields that describe the gravitational interaction are the metric and the affine connection, which is {\it a priori} independent from the metric and has dynamics of its own dictated by the action of the theory. Metric-affine (or Palatini) $f(\cR)$ theories are defined by a Lagrangian of the form
\beq
\mathcal{S}=\frac{1}{2\kappa}\int d^4x \sqrt{-g}f(\cR)+S_m
\eeq
where $\cR$ is the Ricci scalar associated to the Ricci tensor of the independent affine connection $\Gamma$ contracted with the metric $g^{\mu\nu}$, and $S_m$ is the matter action, which can be completely general up to this point. As is well known, metric-affine $f(\cR)$ theories do not propagate extra degrees of freedom and are free of ghosts \cite{Olmo:2011uz}, which can now be easily understood by noting that they lie inside the projective-invariant subclass of Ricci-Based gravity theories \cite{Olmo:2022rhf}, whose projective symmetry protects them from propagating pathological degrees of freedom \cite{BeltranJimenez:2019acz,Jimenez:2020dpn}. For a minimally coupled matter sector (in the sense of \cite{Delhom:2020hkb}) with an equation of state with $\omega=1$, e.g. a massless scalar field, this family of theories predicts a modified Friedmann equation of the form
\begin{equation}
3H_{ f(\cR)}^{2}=\frac{f_{\cR}\left[2 \kappa^{2} \rho+\mathcal{R} f_{\cR}-f\right]}{2\left(f_{\cR}+\frac{f_{\cR \cR}}{2} \frac{\dot{\cR}}{H}\right)^{2}}\,,\label{eq:fRFriedmann}
\end{equation}
where the following relations can be derived from the field equations of the theory (see \eg \cite{Olmo:2011uz})
\begin{align}
&\rho=\frac{\cR f_\cR-2f}{2\kappa},\label{eq:DensityOfR}\\
&\frac{\dot{\cR}}{H}=-12\kappa \rho/(\cR f_{\cR\cR}-f_{\cR}) \ .
\label{eq:RdotH}
\end{align}
{From diffeomorphism invariance of $f(\cR)$ theories, and if the affine connection couples minimally to the matter sector, one can show that the above Friedmann equation and conservation of the stress-energy tensor allow to derive the Raychaudhuri equation, the three of them forming an over determined system, just like in GR. (See appendix \ref{sec:Appendix} for a more explicit derivation.)} {The situation is the same in LQC, mLQC-I, and mLQC-II. In this manuscript we use the Friedmann equation and the conservation law to determine the effective action for mLQC-I and mLQC-II from $f(\cR)$ theories.}
{From diffeomorphism invariance of $f(\cR)$ theories, and if the affine connection couples minimally to the matter sector, one can show that the above Friedmann equation and conservation of the stress-energy tensor allow to derive the Raychaudhuri equation, the three of them forming an over determined system, just like in GR. (See appendix \ref{sec:Appendix} for a more explicit derivation.)} {The situation is the same in LQC, mLQC-I, and mLQC-II. In this manuscript we use the Friedmann equation and the conservation law to determine the effective action for mLQC-I and mLQC-II from $f(\cR)$ theories.}

The above relations \eqref{eq:DensityOfR} and \eqref{eq:RdotH} allow to rewrite the right-hand side of \eqref{eq:fRFriedmann} as a function of $f(\cR)$ and its two derivatives $f_\cR$ and $f_{\cR\cR}$ as
\begin{equation}
3H_{ f(\cR)}^{2}=\frac{f_{\cR}\lr{2\mathcal{R} f_{\cR}-3f}}{2\left(f_{\cR}-3f_{\cR \cR}\lr{\frac{\cR f_\cR-2f}{\cR f_{\cR\cR}-f_{\cR}}}\right)^{2}}\,.\label{eq:fRFriedmannODE}
\end{equation}
To find bouncing cosmologies within the $f(\cR)$ subfamily of metric-affine theories we need to have either $f_\cR=0$ or $2\cR f_\cR-3f=0$ at some value of the affine curvature $\cR_\mathrm{bounce}$ and the quotient of \eqref{eq:fRFriedmannODE} being regular as $\cR$ approaches $\cR_\mathrm{bounce}$. We will proceed with the bounce given by the condition $f_\cR=0$. 

\subsection{Physical conditions for equivalence}\label{sec:PhysCondEquiv}

\begin{table*}
\resizebox{\textwidth}{!}{
    \begin{tabular}{|c|c|c|c|}
    \hline
       & A & B & C\\
      \hline
      \textbf{LQC} & $\sqrt{2\left(\mathcal{R} f_{\cR}-2 f\right)\left(2{R}_{\rm c}-\left(\mathcal{R} f_{\cR}-2 f\right)\right)}$ & $2\sqrt{{R}_{\rm c} f_{\cR}\left(2 \mathcal{R} f_{\cR}-3 f\right)}$ & $1$\\
     \hline
     \textbf{mLQC-I$^-$}  & $\sqrt{2(\cR f_{\cR}-2f)\left(\frac{\rc}{2(1+\gamma^2)}-\cR f_{\cR}+2f\right)}$ & $\sqrt{\frac{{R}_{\rm c}}{1+\gamma^2} f_{\cR}\left(2 \mathcal{R} f_{\cR}-3 f\right)}$ & $\sqrt{\frac{\rc-(1+\gamma^2)(\cR f_\cR-2f)+\sqrt{\rc^2-2(1+\gamma^2)\rc(\cR f_{\cR}-2f)}}{(1+\gamma^2)\left(\rc+2f-\cR f_\cR+\sqrt{\rc^2-2(1+\gamma^2)\rc(\cR f_{\cR}-2f)}\right)}}$\\
     \hline
     \textbf{mLQC-I$^+$}  & $\sqrt{\frac{4 \sqrt{3} \left(\frac{\rc}{2 \left(\gamma ^2+1\right)}-R f_{\cR}+2 f\right)}{\left(\gamma ^2+1\right) \gamma   \kappa ^2 \rc}}$ & $\sqrt{\frac{{R}_{\rm c}}{1+\gamma^2} f_{\cR}\left(2 \mathcal{R} f_{\cR}-3 f\right)}$ & $\sqrt{\frac{\gamma^2\left(\sqrt{\frac{\rc}{2(1+\gamma^2)}}+\sqrt{\frac{\rc}{2(1+\gamma^2)}-\cR f_{\cR}+2f}
\right)}{2\gamma^2+1-\sqrt{\frac{\rc}{2(1+\gamma^2)}-\cR f_{\cR}+2f}
}}$\\
     \hline
     \textbf{mLQC-II}  & $\sqrt{2\left(\mathcal{R} f_{\cR}-2 f\right)\left(2{R}_{\rm c}+\gamma^2\left(\mathcal{R} f_{\cR}-2 f\right)\right)}$ & $2\sqrt{{R}_{\rm c} f_{\cR}\left(2 \mathcal{R} f_{\cR}-3 f\right)}$ & $\sqrt{\frac{\gamma^2\left(1+\sqrt{\gamma^2\frac{\cR f_\cR-2f}{2\rc}+1}\right)}{2\gamma^2+1-\sqrt{\gamma^2\frac{\cR f_\cR-2f}{2\rc}+1}}}$\\
     \hline
    \end{tabular}
    }
    \caption{\label{tab:EqCoef}
This table contains the values of the different factors that define the particular form of the general ODE which enforces equivalence between the evolution of metric-affine $f(\cR)$ theories and the different loop cosmology models that we will deal with.}
\end{table*}

The first condition for evolution equivalence can be formulated as an equivalence of the Hubble rates of both pictures. On the explicitly covariant side, the Hubble rate is a function of the metric-affine $f(\cR)$ Lagrangian, and its first two derivatives, as seen in \eqref{eq:fRFriedmann}. On the other hand, in the loop cosmology side, the Hubble rate can be written in terms of the energy density for each of the models as in Eqs. \eqref{eq:LQCFriedmann} to \eqref{eq:mLQC-IFriedmann}. Using now the relations that exist in the $f(\cR)$ side between energy density, the Lagrangian, and its first derivative \eqref{eq:DensityOfR}; we can enforce the equivalence of Hubble rates by requiring that the $f(\cR)$ Lagrangian solves a 2nd-order nonlinear ODE. This equation will be different for each of the loop cosmology models here considered, but they can all be written in the general form\footnote{This form will still hold for perfect fluids even if $\omega\neq 1$, as can be derived from the corresponding $\omega$-dependent version of Eq. (7). However the specific form of A, B and C will depend on $\omega$.}
{
\begin{equation}\label{eq:GeneralODE}
f_{\cR\cR}=f_\cR \lr{\frac{B C-f_{\cR}A}{B C \cR+2A(\cR f_{\cR}-3f)}}
\end{equation}
where $A$, $B$ and $C$ are functions of $\cR$, $f(\cR)$ and $f_\cR$ which depend on the loop cosmology models under consideration. }We show the particular values of these coefficients for LQC, mLQC-II and the two asymmetric branches of mLQC-I in Table I.

Given a pair of boundary conditions for $f(\cR)$ and $f_\cR$ at some value of $\cR$, this equation yields a unique solution in the interval $[\cR_\mathrm{bounce},0]$ provided that the right hand side of \eqref{eq:GeneralODE} is regular within the whole interval. The required boundary conditions can be found analytically by enforcing physical equivalence at the bounce in the values of the energy density and the acceleration of the scale factor for the two descriptions. { Under the constraint $f_\cR=0$, enforced in order to have a bounce in the $f(\cR)$ description\footnote{There are other bouncing solutions with $f_\cR\neq 0$, but these end up determining the values of $f_{\cR\cR}$ and even higher derivatives at the bounce, so that they are not suited to yield initial conditions for the ODE in \eqref{eq:GeneralODE}.}, these conditions lead to a unique set of allowed boundary conditions: imposing equality of the Hubble function, the bounce density yields the value of the Lagrangian at the bounce, while enforcing equality in the acceleration of the scale factor yields the value of the affine curvature at the bounce, which has the value of $-12 \dot{H}_\textrm{bounce}$ for the three models}. The specific values for each of the models are given in  Table \ref{tab:BounceCond}. With these initial conditions we can, in principle, obtain a solution to the corresponding ODE, which would provide a particular Lagrangian of the $f(\cR)$ type for each of the loop cosmology models. Note that, according to the definitions on Table \ref{tab:EqCoef}, there is no obvious reason to expect that these Lagrangians should bear any relation among them, beyond the fact of laying within the metric-affine $f(\cR)$ subclass. However, as we will see below, their nonperturbative behavior can be parametrised by a subfamily of $f(\cR)$ functions where nonperturbative deviations from the Einstein-Hilbert Lagrangian arise via logarithmic corrections of the form $\ln(\cR/\cR_\mathrm{bounce})$. 

\begin{table}
  \begin{center}
    \begin{tabular}{|c|c|c|c|}
    \hline
      &   \textbf{LQC} & \textbf{mLQC-I} & \textbf{mLQC-II}\\
      \hline
     $\mathcal{R}_\mathrm{b}$ & $-12\rc$ & $-\frac{3(1+2\gamma^2)}{(1+\gamma^2)^2}\rc$ & $-48(1+	\gamma^2)(1+2\gamma^2)\rc$\\
     \hline
     $f(\mathcal{R}_\mathrm{b})$  &$-\rc$ & $-\frac{1}{4(1+\gamma^2)}\rc$ & $-4(1+\gamma^2)\rc$\\
     \hline
    \end{tabular}
    \caption{
This table contains the values of the affine curvature and the effective Lagrangian at the Bounce conditions for the equivalence between the metric-affine $f(\cR)$ description and the effective dynamics of the respective loop cosmology models. They are found by enforcing equivalence of the bounce energy density and the acceleration of the scale factor.}
\label{tab:BounceCond}
  \end{center}
\end{table}

\section{Numerical solutions}\label{sec:NumSol}

Though an analytical solution of the ODEs that encode evolution equivalence for the different models would be ideal, to the best of our knowledge, it is not possible to obtain it. Thus, we need to find a numerical solution that we can later approximate with an analytical ansatz. Even obtaining physically meaningful numerical solutions of these ODEs is not generally a straightforward task, due to the fact that the boundary conditions at the bounce, obtained in section \ref{sec:PhysCondEquiv} from requiring physical consistency of both descriptions at the bounce, are not appropriate for numerical analysis, the reason being that the right hand side of \eqref{eq:GeneralODE} is indeterminate for these initial conditions. Hence, we have to look for boundary conditions elsewhere within the interval $[\mathcal{R}_\textrm{bounce},0]$ and find the appropriate numerical solution by requiring that the bounce that it describes satisfies the physically consistent bounce conditions up to a certain degree of precision.

Given that, in order for the effective $f(\cR)$ Lagrangians to be physically meaningful, they must approximate GR at low curvatures, we can try to fix boundary conditions near $\cR=0$, where we know that $f_\cR$ should be close to unity. Given a value of $\cR$ at which we can fix boundary conditions, $\mathcal{R}_\textrm{bounce}$, the values of $f(\cR)$ and $f_\cR$ at $\mathcal{R}_\textrm{bounce}$ can be tuned so that the numerical solution reproduces the right conditions at the bounce up to the desired accuracy. In that sense, we note that, for any choice of $\mathcal{R}_\textrm{bounce}$ within the interval $[\mathcal{R}_\textrm{bounce},0]$, there is an upper limit to the boundary value of $f_\cR$ above which the numerical solution diverges before reaching the bounce. Below this critical value, the ODEs are well behaved and the conditions for existence and uniqueness of the solution are satisfied across the relevant interval $[\mathcal{R}_\textrm{bounce},0]$. More interestingly, it appears that the physically consistent bounce conditions are best approximated for the numerical solution with a boundary value for $f_\cR$ equal to this critical value. The robustness of these numerical results has been verified by giving boundary values close to the critical value, both above and below, and running an iterative Newton-Raphson method where the goal is to achieve $f_\cR=0$ at the bounce. The iterations show how the boundary value of the derivative converges to the critical value as the number of iterations increase in both LQC and mLQC-II, as well as in mLQC-I$^-$. However, the solutions are extremely sensitive to a small change in the value of the derivative. A deviation in $f_\cR$ on the seventh significant digit at $\mathcal{R}_\textrm{bounce}=10^{-25}$ yields variations in the energy density at the bounce of around the 10\% of its value. Nonetheless, note that it is remarkable that there are solutions which satisfy the appropriate boundary conditions at the bounce to a desired degree of precision while at the same time reducing to GR at low curvatures. This could signal that the Einstein-Hilbert Lagrangian might be an attractor at $\cR=0$ in solution space as, otherwise, solutions satisfying the boundary bounce conditions would not be, in general, close to GR near $\cR=0$.

\begin{figure*}[htb!]
\begin{center}
\includegraphics[width=0.44\textwidth]{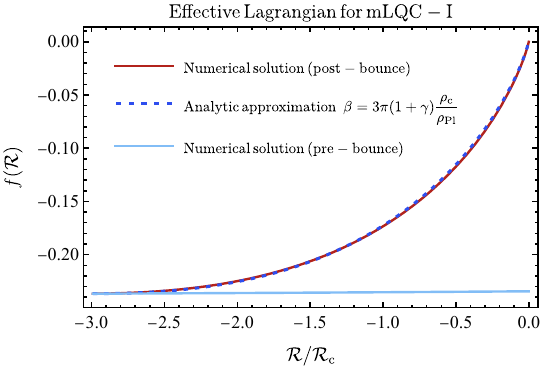}\hspace{1cm}
\includegraphics[width=0.423\textwidth]{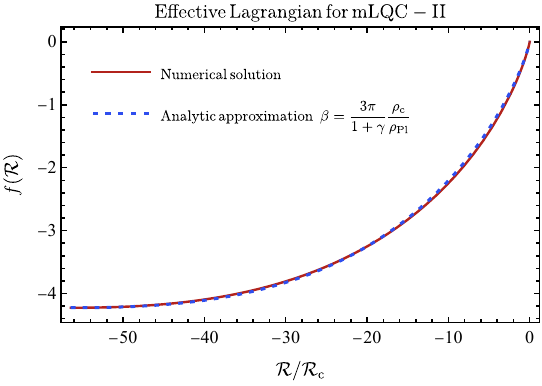}
\caption{Plots of the effective Lagrangian given by the numerical solutions to \eqref{eq:GeneralODE} for mLQC-I and mLQC-II and its analytical approximations given by \eqref{GenericModel} with the parameters in Table \ref{tab:ParamVal} and eqs. \eqref{eq:betaLQC} and \eqref{eq:Relationbetas}. The values of the affine curvature and energy density are given in units of $\rc$ and the bounce curvature for each model. We can see how the pre-bounce branch is described by an almost constant Lagrangian as corresponds to a quasi-de-Sitter phase.}
\label{fig:mLQCI}
\end{center}
\end{figure*}

The method of placing boundary conditions at low curvatures is appropriate for the post-bounce phases where, according to our experience, the gravity Lagrangian must resemble the Einstein-Hilbert action at low curvatures up to a high degree of precision. For LQC and mLQC-II, this procedure completely determines a covariant Lagrangian that describes the full cosmological evolution at the background level, due to the fact that the pre- and post-bounce branches are symmetric within these models. However, this is not true for mLQC-I, for which the pre-bounce branch differs from the post-bounce one in cosmic time.   Thus, though we have experimental data about how the low-curvature regime of the mLQC-I post-bounce evolution looks like, this is not the case for the pre-bounce branch, for which we do not have any available data constraining the behavior of the Lagrangian at low curvatures, even though this phase has been constrained by CMB observations \cite{Agullo:2018wbf,Li:2019qzr,Li:2020mfi,Gomar:2020orw}. Indeed, the pre-bounce branch is asymptotically de Sitter with an asymptotic curvature that is not far from the Planck scale.\footnote{ {{Such an asymmetric evolution is a direct manifestation of the form of the Hamiltonian constraint in mLQC-I, with a similar evidence seen also for a loop quantization of Kantowski-Sachs spacetime \cite{Dadhich:2015ora}. }}}
This makes it impossible to assess the low curvature behavior of the corresponding effective $f(\cR)$ Lagrangian for the pre-bounce phase, as we know that, once the curvature reaches the value of the effective cosmological constant, the curvature-dependent terms of the Lagrangian become irrelevant as compared to the effective cosmological constant, thus leading to a high degree of arbitrariness. Thus, we can only hope for unveiling the form of the Lagrangian from scales close to the effective cosmological constant to the bounce curvature.

Even so, we have been able to find numerical solutions for the pre-bounce phase of mLQC-I which match the value of the energy density, Lagrangian, and first derivative of the post-bounce phase. Both pre-bounce and post-bounce numerical solutions are plotted in figure \ref{fig:mLQCI}. We see that the solutions are almost constant, both for the Lagrangian and the energy density. However, the values of these numerical solutions do not correspond to the correct values of the asymptotically de Sitter phase, which are $f_{\mathrm{asymp.}}(\cR)= -0.297$ and $\rho_\Lambda=0.314\rhob^\mathrm{I}$. Our investigations show that it is difficult to approach more precise values of the asymptotic energy density by changing the boundary conditions at the bounce, due to the sensitivity of the ODE to small changes in the boundary conditions.

\section{Analytical approximations}\label{sec:Analytics}

Following \cite{Olmo:2008nf}, we now approximate the numerical solutions that describe LQC, mLQC-II and the post-bounce phase of mLQC-I. Our aim is to see whether the quantum gravitational  non-perturbative behavior of the different models can be fitted into the same family of metric-affine Lagrangians. Hence, we proceed by generalizing the analytical approximation to LQC introduced in \cite{Olmo:2008nf} to a three parameter family of metric-affine $f(\cR)$ Lagrangians of the form 
\begin{equation}\label{GenericModel}
f(\cR)=\xi \cR\left[1+\frac{(\cR-\alpha \rc)^2}{\beta \alpha^2 \rc^2}-\frac{1}{2}\ln\left(\frac{\cR^2}{\alpha^2 \rc^2}\right)\right] .
\end{equation}
Here the nonperturbative effects near the bounce can be mostly accounted for by a logarithmic correction. The parameter  $\alpha$ determines the value of the affine curvature at the bounce in units of $\rc$, and $\xi$ determines the value of the Lagrangian at the bounce. $\beta$ is a free parameter controlling the perturbative information of the effective action (namely its polynomial piece) that can be used to fit the numerical solution. In order for $f_\cR$ to have only one root on the interval $[0,\mathcal{R}_\textrm{bounce}]$, it is necessary that $\beta\geq2$. By requiring that the bounce curvature and value for the Lagrangian are consistent with the respective models, the values of $\alpha$ and $\xi$ are completely determined for each case (see table \ref{tab:ParamVal}), leaving only $\beta$ to fit the numerical results. As we will see later, there is an unexpected relation between the values of $\beta$ obtained by doing independent fits for each of the models in terms of the Barbero-Immirzi parameter given by
\begin{equation}\label{eq:Relationbetas}
\beta_{\textrm{I}}=(1+\gamma)\beta_{\textrm{LQC}}\qquad\text{and}\qquad \beta_{\textrm{II}}=\frac{\beta_{\textrm{LQC}}}{(1+\gamma)}.
\end{equation}

\begin{figure*}[htb!]
\centering
\includegraphics[width=0.45\textwidth]{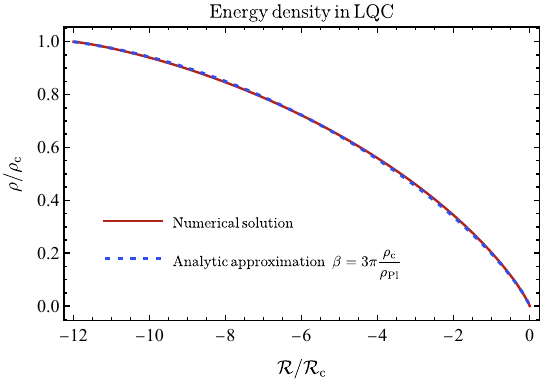}\\
\includegraphics[width=0.45\textwidth]{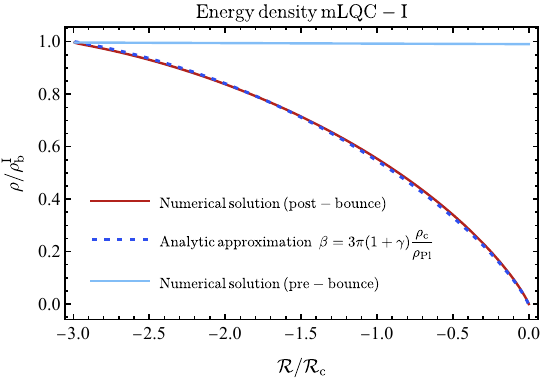}\hspace{1cm}
\includegraphics[width=0.45\textwidth]{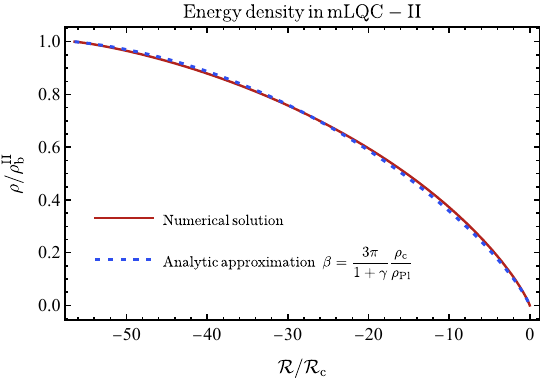}

\caption{Plot of the energy density in units of the corresponding bounce densities for the numerical solution to \eqref{eq:GeneralODE} and the analytical approximation of each model given by \eqref{GenericModel} with the parameters in Table \ref{tab:ParamVal} and eqs. \eqref{eq:betaLQC} and \eqref{eq:Relationbetas}. The top plot is LQC, the bottom left is mLQC-I and the bottom right is mLQC-II. order of the models is standard LQC, post-bounce branch of mLQC-I, and mLQC-II. Note that for LQC we have used a the value for $\beta$ in terms of fundamental LQC parameters  \eqref{eq:betaLQC} $(\approx3.85)$, which coincides with our best-fit value, instead of the one used in \cite{Olmo:2008nf} $(\approx 3.76)$, although both results are practically indistinguishable at the observable level. The mLQC-I analytic approximation captures the non-perturbative behavior of the energy density at high curvatures very accurately and, although some deviation can be found at intermediate scales, this can be overcome by adding fine tuned polynomial corrections at the $1\%$ level if desired. The pre-bounce branch is almost constant as corresponding to an asymptotically de-Sitter phase, although its value does not correspond to the expected one $\rho_\Lambda\approx 0.3\rho_\textrm{b}^\textrm{I}$. We can seee that the behavior of the energy density is accurately descirbed by the analytic approximation at all scales, particularly near the bounce, for the post-bounce phase of mLQC-I as well as for all regimes of LQC and mLQC-II.}
\label{fig:DensityPlot}
\end{figure*}

\begin{table}
  \begin{center}
    \begin{tabular}{|c|c|c|}
    \hline
      &  $\alpha$ & $\xi$ \\
       \hline
       \textbf{LQC} & $-12$ & $-\frac{1}{\alpha_\textrm{LQC}}$\\
        \hline
       \textbf{mLQC-I$^-$}  & $-\frac{3 (1+2\gamma^2)}{(1+\gamma^2)^2}$ & -$\frac{1}{4(1+\gamma^2)\alpha_{\textrm{I}}}$\\
       \hline
       \textbf{mLQC-II}  & $-48(1+2\gamma^2)(1+\gamma^2)$ & $-\frac{4(1+\gamma^2)}{\alpha_{\textrm{II}}}$\\
     \hline
    \end{tabular}
    \caption{This table contains the values of $\alpha$ and $\xi$ that have to be chosen in the general Lagrangian \eqref{GenericModel} so that it yields the correct values of the affine curvature and the effective Lagrangian at the bounce for each of the loop cosmology models. These conditions determine two of the three parameters of the family. The third parameter can be found by fiting the numerical solutions for each of the models, and related to fundamental constants from the underlying quantum dynamics by \eqref{eq:betaLQC} and \eqref{eq:Relationbetas}.}
    \label{tab:ParamVal}
 
 \end{center}
\end{table}

Also, note that the value of $\beta$ is (inversely) correlated to the bounce density of each model. The best fit values for $\beta$, obtained by minimising the area between the numerical solution and the analytical approximation, are given by
\begin{align}\label{betafits}
&\beta_\textrm{LQC}=3.84 \hspace{1cm} \beta_{\textrm{I}}^{-}=4.77 \hspace{1cm} \beta_{\textrm{II}}=3.10.
\end{align}
where $\beta_{\textrm{I}}^{-}$ refers to the post-bounce phase of mLQC-I. This values satisfy the relation \eqref{eq:Relationbetas} to the reported precision. It is also interesting to note that the above value for $\beta_{\textrm{LQC}}$ is also approximated by 
\begin{equation}\label{eq:betaLQC}
\beta_{\textrm{LQC}}=3\pi \frac{\rhoc}{\rho_{\textrm{Pl}}}.    
\end{equation}

\begin{figure}
\begin{center}
\includegraphics[width=0.5\textwidth]{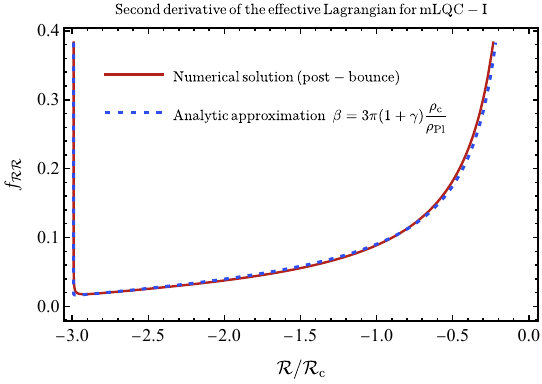}
\caption{Second derivative of the efective Lagrangian given by the numerical solutions to \eqref{eq:GeneralODE} for mLQC-I  and the analytical approximation given by \eqref{GenericModelSquareRoot} with the parameters in Table \ref{tab:ParamVal} and eqs. \eqref{eq:betaLQC} and \eqref{eq:Relationbetas}. The square-root term in \eqref{GenericModelSquareRoot} is necessary to account for the non-perturvative behavior of the second derivative in the Lagrangian, and it does not contribute to the energy at the bounce, so that the behavior of the energy density is still well described by the universal Logarithmic correction in \eqref{eq:GeneralODE}. The square-root term in \eqref{GenericModelSquareRoot} adds a constant term to the Lagrangian that causes the small deviations at lower curvatures seen in fig. \ref{fig:DensityPlot}. These deviations can be solved by adding a small polynomial correction to \eqref{GenericModelSquareRoot}.}
\label{fig:2der}
\end{center}
\end{figure}

Indeed, the best fits obtained with the $\beta$ for each model as a free parameter and those obtained by the above value for $\beta_{\textrm{LQC}}$ and  the relations \eqref{eq:Relationbetas} to determine all the $\beta$ parameters in $\beta_{\textrm{II}}$ in terms of fundamental constants of LQC lead to analytic approximations that are virtually indistinguishable from one another, see figure  \ref{fig:DensityPlot}. As we can see, the non-perturbative behavior is matched by the analytic approximations to a high degree of accuracy, finding deviations only at the percent level in mLQC-I$^-$ and -II. However, if one looks close to the mLQC-I bounce ($\cR=0.999 \cR_{\textrm{b}}$), it is possible to see that the energy density does not approach the bounce with the correct asymptotic behavior. This can be understood by noticing that while the non-perturbative behavior of the Lagrangian at the bounce is perfectly fit in LQC and mLQC-II, the second derivative of the effective Lagragian has a divergence at the bounce for the numerical solution in the mLQC-I case, which is not accounted for by the logarithmic corrections. By studying the asymptotic behavior of the numerical solution, it can be seen that this divergence corresponds to a square-root-like asymptotic behavior of the first derivative, so that the second derivative diverges when approaching the bounce as $(\alpha\rc-R)^{-1/2}$. This can be corrected by modifying the above family with a square root term, so that the mLQC-I$^-$ branch is approximated by
\begin{equation}\label{GenericModelSquareRoot}
\begin{split}
f(\cR)=&\xi \cR\left[1+\frac{(\cR-\alpha \rc)^2}{\beta \alpha^2 \rc^2}-\frac{1}{2}\ln\left(\frac{\cR^2}{\alpha^2 \rc^2}\right)\right]+\frac{\gamma^2}{(2\pi)^3}\sqrt{\frac{(\alpha \rc-\cR)^3}{\alpha \rc}}
\end{split}
\end{equation}
which perfectly matches the behavior of the numerical solution, including the divergenve at the bounce of its second derivative (see fig. \ref{fig:2der}). Although this correction seems to drive the effective metric-affine description of mLQC-I away from the same family of theories as LQC or mLQC-II, we note that this is not the case if one focuses only on the correspondence of the relevant physical observables near the bounce. In fact, the contribution of the square-root correction to the energy density of the effective metric-affine description is given by
\begin{equation}
\frac{{\gamma^2}}{32 \pi^3  \kappa} (\cR-4 \alpha  {\rc}) \sqrt{\frac{\alpha \rc-\cR}{\alpha\rc}},
\end{equation}
which describes with slightly more accuracy the asymptotic behavior of $\rho$ as we approach the bounce but which vanishes at the bounce, although it introduces a negative contribution to the energy density at vanishing curvature given by $-\gamma^2\alpha\rc/ 8\pi^3\kappa$. By taking the appropriate limits, one can also check that the contribution of the correction to the Hubble factor vanishes both, at the bounce, and at low curvatures, while the global fit is as good as the one without the square root correction. This shows that a logarithmic correction is universal in all three loop cosmology models in accounting for the non-perturbative behavior of the energy density as the bounce is approached.

We conclude that the family of theories given in \eqref{GenericModel} provides an accurate description of the full evolution of LQC and mLQC-II, as well as the post-bounce phase of mLQC-I.  Let us, however, elaborate on the need of introducing the square-root correction in \eqref{GenericModelSquareRoot} to fit the behaviour of the second derivative of the Lagrangian with respect to the curvature for mLQC-I. A possible explanation for the physical need of this correction would be that, due to the fact that mLQC-I has an asymmetric bounce, then both branches have to be described by different effective $f(\cR)$ models, which opens the door to have a non-smooth Lagrangian at the bounce, even if this does not permeate all physical observables. Although this seems natural from the loop cosmology side, the reader should notice that the ODE for the mLQC-I post-bounce branch does not have any encoded information about the pre-bounce branch. Therefore, it is rather surprising that, precisely for this model but not for the symmetric ones, the smoothness of the Lagrangian is lost at the bounce. On the other hand, the fact that this lack of smoothness does not affect the relevant physical observables, in principle, allows for a physically well behaved transition from one branch to the other.

\section{Outlook}

In this work we have established a correspondence between the evolution of different loop cosmology models and that of metric-affine $f(\cR)$ theories at the background level, expanding previous results and proving that the background dynamics of different loop cosmologies appears to be generically indistinguishable from that of some diffeomorphism invariant theory. Furthermore, we have shown that the evolution of the energy density with the curvature can be accurately described for the three models of loop cosmologies analyzed here by a single family of $f(\cR)$ theories.  This is so despite the fact that one of the loop cosmology models has an asymmetric bounce, while the other two have symmetric ones. 

Structural differences in the effective Lagrangian of the symmetric and asymmetric bounce models only become relevant in the second derivative of the Lagrangian with respect to the affine curvature, which for the symmetric bounce models is well captured by the family \eqref{GenericModel}, but requires an extra term given in \eqref{GenericModelSquareRoot} for  mLQC-I in order to faithfully track the divergent behavior of $f_{\cR\cR}$ as the bounce is approached. We ignore if  this divergence of $f_{\cR\cR}$ in mLQC-I is just an accident or a generic signature of asymmetric models. 
In any case, the lack of analyticity in $f_{\cR\cR}$ 
has no effect on physical aspects such as the energy density or the acceleration of the expansion factor at the bounce, which are the key observables used to construct our numerical solutions starting with initial data close to GR at low curvatures. In fact, the square root correction of mLQC-I in \eqref{GenericModelSquareRoot} does not contribute to the energy density or the Hubble rate at the bounce, so that the non-perturbative behavior of the relevant physical observables is not affected by it. Furthermore, the near-bounce behavior of the energy density is universally well accounted for by a logarithmic correction as in \eqref{GenericModel}, as can be seen in fig. \ref{fig:DensityPlot}. A limitation of our analysis is that for the pre-bounce branch of mLQC-I we have been unable to recover the asymptotic value of the energy density in the emergent Planckian phase.

Finally, let us comment on the fact that the effective dynamics of the three loop cosmology models can be embedded into the same family of metric-affine $f(\cR)$ Lagrangians. This is an a priori unexpected result. In fact, the proposed family of metric-affine $f(\cR)$ theories does not encode any information about the common origin of the three models and the ambiguities that generate them so that, in principle, there was no guarantee that they could fit within a one-parameter family of Lagrangian theories. It is also remarkable the fact that the only free parameter of the family that can be used to fit the behavior of the energy density satisfies a simple relation between its best fit values for the three models given in \eqref{eq:Relationbetas}, and even more remarkable that the values of this parameter and these relations can be written in terms of fundamental constants that originate in the underlying quantum description. These numerical coincidences call for a closer scrutiny in order to understand whether they are mere chance or if there exists some underlying reason for this to occur. 

In this sense, to understand the significance of these results concerning the universality of the three effective descriptions, one should assess the generality of the logarithmic term in nonsingular metric-affine cosmologies. It is important to clarify if it is a general feature of models with an upper bound on the energy density or if it is somehow characteristic of the quantum effects of loop cosmologies.   Another important task consists on exploring other scenarios with different symmetry-reduction patterns in phase space, such as black hole like space-times, where LQG-related quantum corrections can also lead to singularity resolution and upper bounds on the space-time curvature \cite{Ashtekar:2018lag}.

\appendix

\section{Field equations of $f(\cR)$ theories and Raychaudhuri equation}\label{sec:Appendix}

The field equations of Palatini $f(\cR)$ theories can be derived from the general expression of Ricci-Based Gravity theories \cite{Olmo:2022rhf}, which take the generic form
\begin{equation}\label{eq:RBGs}
{G^\mu}_\nu(h)=\frac{\kappa}{|\Omega|^{1/2}}\left[{T^\mu}_\nu-\delta^\mu_\nu \left(\mathcal{L}_G+\frac{T}{2}\right)\right] \ ,
\end{equation}
where $\mathcal{L}_G$ represents the gravity Lagrangian, $\mathcal{L}_G=f(\cR)/2\kappa$ in our case, and $|\Omega|$ represents the determinant of the deformation matrix that relates the connection compatible metric $h_{\mu\nu}$ and the space-time metric $g_{\mu\nu}$, namely, $h_{\mu\nu}={\Omega_\mu}^\alpha g_{\alpha\nu}$. In the $f(\cR)$ case, we have ${\Omega_\mu}^\alpha=f_\cR{\delta_\mu}^\alpha$, thus leading to $h_{\mu\nu}=f_\cR g_{\mu\nu}$ and $|\Omega|^{1/2}=f_\cR^2$. Note that the upper index on the left-hand side of (\ref{eq:RBGs}) is raised with $h^{\mu\nu}$ whereas on the right-hand side we use $g^{\mu\nu}$. 

The Friedman equation is obtained from the ${G^t}_t(h)$ component using a line element of the form $ds_h^2=f_\cR(-dt^2+a^2(t)d\vec{x}^2)$, which yields
\begin{equation}
    {G^t}_t(h)\equiv -\frac{3}{f_\cR}\left(H+\frac{1}{2}\frac{\dot f_\cR}{f_\cR}\right)^2=-\frac{\kappa}{f_\cR^2}\left(\rho+\mathcal{M}\right) \ ,
\end{equation}
where $\mathcal{M}\equiv f(\cR)/2\kappa+T/2$. The final expression for $H^2$ is obtained by noting that the algebraic relation $\cR f_\cR-2f=\kappa T$ leads to 
\begin{equation}
\dot \cR=\frac{\kappa \dot T}{\cR f_{\cR\cR}-f_\cR} \ .
\end{equation}
This can be used to write $\dot f_\cR=f_{\cR\cR}\dot \cR$, and together with the conservation equation for a perfect fluid with $P=\omega \rho$, {where $\omega$ is a constant}, we find that $\dot T=3H(1+3\omega)\rho$. Putting all together, when $\omega=1$ we find the expression used in Eq.(\ref{eq:fRFriedmannODE}) for a massless scalar field. 

Given the formal similitude between the RBG equations and those of GR, the Raychaudhuri equation for $f(\cR)$ follows from the same manipulations as in the GR case. The combination ${G^x}_x-\frac{1}{3}{G^t}_t$ leads to 
\begin{eqnarray}
    {G^x}_x-\frac{1}{3}{G^t}_t&\equiv& \frac{1}{f_\cR}\left[\left(\frac{\dot f_\cR}{f_\cR}\right)^2-\frac{2\ddot a}{a}-\left(H\frac{\dot f_\cR}{f_\cR}+\frac{\ddot f_\cR}{f_\cR}\right)\right]\nonumber \\
    &=& \frac{\kappa}{3f_\cR^2}\left[(\rho+3P)-2\mathcal{M}\right] \ ,
\end{eqnarray}
which clearly boils down to the standard GR result when $f(\cR)=\cR$. 

\begin{figure*}[htb!]
\begin{center}
\hspace{-0.5cm}
\includegraphics[width=0.45\textwidth]{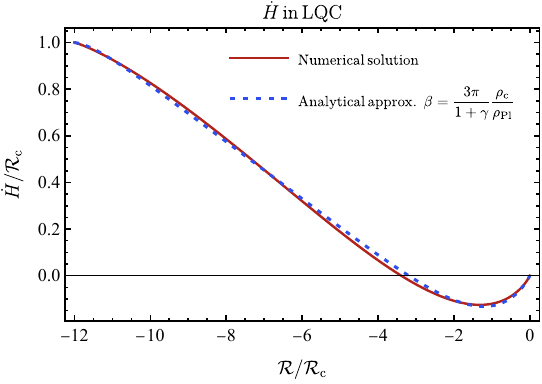}\\
\includegraphics[width=0.45\textwidth]{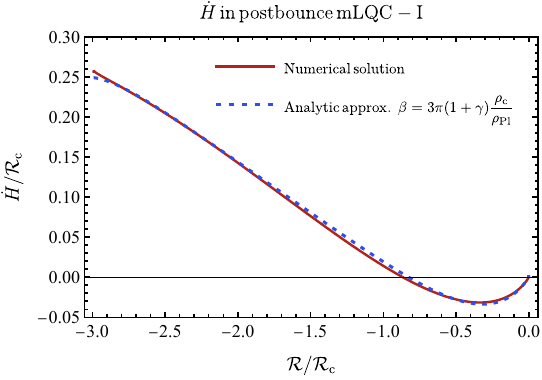}
\includegraphics[width=0.45\textwidth]{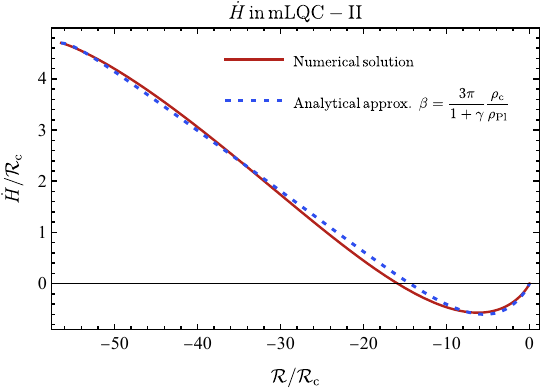}
\caption{$\dot{H}$ as a function of the affine curvature for the analytic approximations in \eqref{GenericModel} and \eqref{GenericModelSquareRoot} and the corresponding numerical solutions. We see that the approximations are quite faithful through all curvature ranges.}
\label{fig:Hdot}
\end{center}
\end{figure*}

{Diffeomorphism invariance guarantees $\nabla_\mu^{(g)} {T^\mu}_\nu =0$, and using the explicit relation between $\nabla_\mu^{(g)}$ and $\nabla_\mu^{(h)}$, it is straightforward to see that the Bianchi identity $\nabla_\mu^{(h)} {G^\mu}_\nu (h)=0$ also applies on the right-hand side of (\ref{eq:RBGs}).} From this identity we can find the relation ${G^x}_x={G^t}_t+\frac{1}{\lambda}{\dot G^t}_t$, where 
\begin{equation}
\lambda= 3H+\frac{3}{2}\frac{\dot f_\cR}{f_\cR}    \ .
\end{equation}
This proves that, like in GR, in $f(\cR)$ theories the corresponding Friedman and Raychaudhuri equations together with the conservation equation $\nabla_\mu^{(g)} {T^\mu}_\nu =0$ define an over determined system in which only two of those equations are really independent. { In the main body of this manuscript we have used the Friedmann equation and the conservation law in $f(\cR)$ theories to obtain the effective action for mLQC-I and mLQC-II.} {For completeness, we plot in figure \ref{fig:Hdot} $\dot{H}$ as a function of the affine curvature for the numerical solutions and analytic approximations, finding also good agreement between both.}

\acknowledgments
We would like to thank Beatriz Elizaga-Navascu\'es for useful comments on the manuscript. AD is supported by NSF grants PHY-1903799, PHY-2206557 and funds from the Hearne Institute for Theoretical Physics. PS is supported by NSF grant PHY-2110207. This work is also supported by the Spanish Grant PID2020-116567GB- C21 funded by MCIN/AEI/10.13039/501100011033, the project PROMETEO/2020/079 (Generalitat Valenciana), and by the European Union’s Horizon 2020 research and innovation program under the H2020-MSCA-RISE-2017 Grant No. FunFiCO-777740.

\bibliographystyle{JHEP}
\bibliography{Bibliography.bib}

\end{document}